\documentclass{article}

\usepackage{arxiv}

\usepackage[utf8]{inputenc} 
\usepackage[T1]{fontenc}    
\usepackage{hyperref}       
\usepackage{url}            
\usepackage{booktabs}       
\usepackage{amsfonts}       
\usepackage{nicefrac}       
\usepackage{microtype}      
\usepackage{lipsum}
\usepackage{graphicx}

\title{Exploring complex networks with the \texttt{ICON} R package}

\author{
    Raoul R. Wadhwa
   \\
    Cleveland Clinic Lerner College of Medicine \\
    Case Western Reserve University \\
  Cleveland, OH 44195 \\
  \texttt{\href{mailto:raoul.wadhwa@case.edu}{\nolinkurl{raoul.wadhwa@case.edu}}} \\
   \And
    Jacob G. Scott
   \\
    Translational Hematology \& Oncology Research \\
    Lerner Research Institute, Cleveland Clinic \\
  Cleveland, OH 44195 \\
  \texttt{\href{mailto:ScottJ10@ccf.org}{\nolinkurl{ScottJ10@ccf.org}}} \\
  }

\usepackage{color}
\usepackage{fancyvrb}

\DefineVerbatimEnvironment{Highlighting}{Verbatim}{commandchars=\\\{\}}
\usepackage{framed}
\definecolor{shadecolor}{RGB}{248,248,248}
\newenvironment{Shaded}{\begin{snugshade}}{\end{snugshade}}

\newcommand{\CommentTok}[1]{\textcolor[rgb]{0.56,0.35,0.01}{\textit{#1}}}

\newcommand{\DataTypeTok}[1]{\textcolor[rgb]{0.13,0.29,0.53}{#1}}
\newcommand{\DecValTok}[1]{\textcolor[rgb]{0.00,0.00,0.81}{#1}}

\newcommand{\FloatTok}[1]{\textcolor[rgb]{0.00,0.00,0.81}{#1}}

\newcommand{\KeywordTok}[1]{\textcolor[rgb]{0.13,0.29,0.53}{\textbf{#1}}}
\newcommand{\NormalTok}[1]{#1}
\newcommand{\OperatorTok}[1]{\textcolor[rgb]{0.81,0.36,0.00}{\textbf{#1}}}
\newcommand{\OtherTok}[1]{\textcolor[rgb]{0.56,0.35,0.01}{#1}}

\newcommand{\StringTok}[1]{\textcolor[rgb]{0.31,0.60,0.02}{#1}}

\begin{document}
\maketitle

\def\tightlist{}

\begin{abstract}
We introduce \texttt{ICON}, an R package that contains 1075 complex network datasets in a standard edgelist format.
All provided datasets have associated citations and have been indexed by the Colorado Index of Complex Networks - also referred to as ICON. 
In addition to supplying a large and diverse corpus of useful real-world networks, \texttt{ICON} also implements an S3 generic to work with the \texttt{network} and \texttt{ggnetwork} R packages for network analysis and visualization, respectively. 
Sample code in this report also demonstrates how \texttt{ICON} can be used in conjunction with the \texttt{igraph} package.
Currently, the Comprehensive R Archive Network hosts \texttt{ICON} v\(0.4.0\). We hope that \texttt{ICON} will serve as a standard corpus for complex network research and prevent redundant work that would be otherwise necessary by individual research groups. 
The open source code for \texttt{ICON} and for this reproducible report can be found at \url{https://github.com/rrrlw/ICON}.
\end{abstract}


\hypertarget{introduction}{%
\section{Introduction}\label{introduction}}

Empirical study of complex networks requires real-world data to validate theoretical results. 
A large, diverse corpus of networks often proves useful given the many shapes and sizes that complex networks
assume.\textsuperscript{1,2} 
To our knowledge, the \href{https://icon.colorado.edu}{Colorado Index of Complex Networks (ICON)} hosts the largest curated index of real-world complex networks, with metadata and links to over 5,000 networks as of this writing.\textsuperscript{3} 
However, heterogeneity in data format, access, and availability limit how easily users can take advantage of this incredible resource.
A central repository containing a large corpus of ICON-indexed networks in standard format would thus provide a useful service for network science researchers, who would avoid the tedious task of data format conversion prior to analysis.
Here, we introduce the \texttt{ICON} R package as such a solution, providing a large and diverse corpus of real-world networks and tools to work with existing network analysis and visualization R packages.

\hypertarget{implementation-details}{%
\section{Implementation details}\label{implementation-details}}

\texttt{ICON} v\(0.4.0\) is a package for the R programming language hosted on the Comprehensive R Archive Network (CRAN).\textsuperscript{4}
It strongly depends on the \texttt{network} (\(\geq 1.16\)) and \texttt{utils} (\(\geq 3.6.1\)) R packages and weakly depends on the \texttt{covr} (\(\geq 3.5\)), \texttt{ggnetwork} (\(\geq 0.5\)), \texttt{ggplot2} (\(\geq 3.3\)), \texttt{knitr} (\(\geq 1.30\)), \texttt{testthat} (\(\geq 2.3\)), and \texttt{rmarkdown} (\(\geq 2.4\)) R packages.\textsuperscript{4--11} 
Throughout the development process, the \texttt{devtools} R package was heavily used.\textsuperscript{12}
This report is fully reproducible with code (GPL-3 license) available at \url{https://github.com/rrrlw/ICON}. Full reproduction will require installation of the strong and weak dependencies listed above.

\hypertarget{installation}{%
\subsection{Installation}\label{installation}}

The stable version of \texttt{ICON} (currently v\(0.4.0\)) can be downloaded from CRAN, while the development version can be downloaded from the package's \href{https://github.com/rrrlw/ICON}{GitHub repository} using the \texttt{remotes} package.\textsuperscript{13} 
Both options are demonstrated in the following code chunk.

\begin{Shaded}
\begin{Highlighting}[]
\CommentTok{# install stable version from CRAN}
\KeywordTok{install.packages}\NormalTok{(}\StringTok{"ICON"}\NormalTok{)}

\CommentTok{# install development version from GitHub}
\NormalTok{remotes}\OperatorTok{::}\KeywordTok{install_github}\NormalTok{(}\StringTok{"rrrlw/ICON"}\NormalTok{, }\DataTypeTok{build_vignettes =} \OtherTok{TRUE}\NormalTok{)}
\end{Highlighting}
\end{Shaded}

\hypertarget{complex-network-datasets}{%
\subsection{Complex network datasets}\label{complex-network-datasets}}

Currently, \texttt{ICON} provides 1075 complex network edge lists. The largest network, named \texttt{amazon\_copurchase}, consists of $3,387,388$
edges. 
Due to the large volume of data and CRAN package size limits, all of \texttt{ICON}'s networks cannot be downloaded to a local machine upon installation and loaded with \texttt{utils::data}.
Instead, the package's GitHub repository contains a directory named
\texttt{data-host}, which \texttt{ICON::get\_data} accesses to download networks named by the user. 
After successful download, \texttt{ICON::get\_data} loads these networks into the user's environment of choice (default: \texttt{.GlobalEnv}) and cleans any intermediate artifacts. 
To avoid dependence on an internet connection, users can save and access individual networks in RDS format (binary; \texttt{.rds} extension; via \texttt{base::saveRDS} and \texttt{base::readRDS}) or do the same for a set of networks in RData/RDa format (binary; \texttt{.RData} or \texttt{.rda} extension; via \texttt{base::save} and \texttt{base::load}). 
An obvious deficiency of this system is the inability to take advantage of automated data documentation and checking tools, such as \texttt{roxygen2}.\textsuperscript{14} 
However, the \texttt{ICON::ICON\_data} dataset provides the necessary documentation for \texttt{ICON} users and implements a sufficient and slightly soporific checking system for the package authors.

Although providing standardized data format avoids redundant work, an important processing step being completed by a single party (package authors) opens the door to inaccuracies. 
It befits us to simply counter this limitation with \texttt{ICON}'s status as free, open-source software (FOSS), which offers every user the opportunity to inspect, question, and correct all aspects of \texttt{ICON}. 
The \texttt{data-raw} directory in \texttt{ICON}'s GitHub repository follows Wickham's (2015) advice and contains: (1) the original raw data acquired directly from the source indexed by the ICON website; (2) the R code that converts each raw dataset into a data frame comprised of an edge list and potential edge attributes; and (3) the R code saving the resulting data frame as an RDA file in the aforementioned \texttt{data-host} directory. We hope that this not only offers, but indeed encourages, \texttt{ICON} users to confirm dataset accuracy.

Note that to minimize unnecessary package elements, \texttt{ICON}'s
\texttt{.Rbuildignore} contains \texttt{data-host} and
\texttt{data-raw}. However, for reproducibility and documentation,
\texttt{ICON}'s GitHub repository provides public access to both
directories.

We will now look at sample code to acquire complex network datasets using \texttt{ICON}. To do so, we must load the library in the R session and load the \texttt{ICON\_data} dataset, which contains relevant
complex network metadata.
The metadata can be explored in the package documentation with \texttt{?ICON\_data}; in this report, we will focus only on the essentials, starting with the following code chunk.

\begin{Shaded}
\begin{Highlighting}[]
\CommentTok{# load library}
\KeywordTok{library}\NormalTok{(}\StringTok{"ICON"}\NormalTok{)}

\CommentTok{# load metadata}
\CommentTok{# explore this data frame to figure out which networks suit your needs}
\KeywordTok{data}\NormalTok{(}\StringTok{"ICON_data"}\NormalTok{)}

\CommentTok{# peek at the first few and last few packages available to download}
\KeywordTok{head}\NormalTok{(ICON_data}\OperatorTok{$}\NormalTok{Var_name, }\DataTypeTok{n =} \DecValTok{3}\NormalTok{)}
\end{Highlighting}
\end{Shaded}

\begin{verbatim}
## [1] "aishihik_intensity"  "aishihik_prevalence" "amazon_copurchase"
\end{verbatim}

\begin{Shaded}
\begin{Highlighting}[]
\KeywordTok{tail}\NormalTok{(ICON_data}\OperatorTok{$}\NormalTok{Var_name, }\DataTypeTok{n =} \DecValTok{3}\NormalTok{)}
\end{Highlighting}
\end{Shaded}

\begin{verbatim}
## [1] "wordadj_french"   "wordadj_japanese" "wordadj_spanish"
\end{verbatim}

We first try downloading a single dataset with \texttt{ICON::get\_data}
and peeking at its contents. Once this succeeds, we confidently download
multiple datasets.

\begin{Shaded}
\begin{Highlighting}[]
\CommentTok{# download single dataset named in previous code chunk output}
\CommentTok{# could also use `get_data(ICON_data$Var_name[1])` to same effect}
\KeywordTok{get_data}\NormalTok{(}\StringTok{"aishihik_intensity"}\NormalTok{)}
\end{Highlighting}
\end{Shaded}

\begin{verbatim}
## DATASET(S) aishihik_intensity LOADED
\end{verbatim}

\begin{Shaded}
\begin{Highlighting}[]
\CommentTok{# look at the structure of the complex network}
\KeywordTok{str}\NormalTok{(aishihik_intensity)}
\end{Highlighting}
\end{Shaded}

\begin{verbatim}
## Classes 'ICON' and 'data.frame': 78 obs. of  3 variables:
##  $ Fish     : chr  "1" "1" "1" "1" ...
##  $ Parasite : chr  "V1" "V9" "V16" "V22" ...
##  $ Intensity: num  5.8 7 3 1 7.2 ...
\end{verbatim}

\begin{Shaded}
\begin{Highlighting}[]
\CommentTok{# confirm that metadata reflects the correct number of edges}
\NormalTok{(ICON_data}\OperatorTok{$}\NormalTok{Edges[}\DecValTok{1}\NormalTok{] }\OperatorTok{==}\StringTok{ }\KeywordTok{nrow}\NormalTok{(aishihik_intensity))}
\end{Highlighting}
\end{Shaded}

\begin{verbatim}
## [1] TRUE
\end{verbatim}

\begin{Shaded}
\begin{Highlighting}[]
\CommentTok{# look at the first few rows; for all ICON datasets:}
\CommentTok{#   columns 1 and 2 = nodes that define the edge}
\CommentTok{#   columns 3 and beyond = edge attributes (e.g. weight)}
\KeywordTok{head}\NormalTok{(aishihik_intensity)}
\end{Highlighting}
\end{Shaded}

\begin{verbatim}
##   Fish Parasite Intensity
## 1    1       V1       5.8
## 2    1       V9       7.0
## 3    1      V16       3.0
## 4    1      V22       1.0
## 5    2       V3       7.2
## 6    2       V8      65.8
\end{verbatim}

\begin{Shaded}
\begin{Highlighting}[]
\CommentTok{# download multiple datasets}
\KeywordTok{get_data}\NormalTok{(}\KeywordTok{c}\NormalTok{(}\StringTok{"wordadj_japanese"}\NormalTok{, }\StringTok{"wordadj_french"}\NormalTok{))}
\end{Highlighting}
\end{Shaded}

\begin{verbatim}
## DATASET(S) wordadj_japanese wordadj_french LOADED
\end{verbatim}

\begin{Shaded}
\begin{Highlighting}[]
\CommentTok{# confirm downloads by looking at internal structure}
\KeywordTok{str}\NormalTok{(wordadj_japanese)}
\end{Highlighting}
\end{Shaded}

\begin{verbatim}
## Classes 'ICON' and 'data.frame': 8300 obs. of  2 variables:
##  $ From: chr  "2" "3" "3" "3" ...
##  $ To  : chr  "3" "4" "5" "9" ...
\end{verbatim}

\begin{Shaded}
\begin{Highlighting}[]
\KeywordTok{str}\NormalTok{(wordadj_french)}
\end{Highlighting}
\end{Shaded}

\begin{verbatim}
## Classes 'ICON' and 'data.frame': 24295 obs. of  2 variables:
##  $ From: chr  "1" "1" "1" "1" ...
##  $ To  : chr  "2" "5" "7" "30" ...
\end{verbatim}

A keen reader might observe that all of \texttt{ICON}'s datasets could
be downloaded with \texttt{get\_data(ICON\_data\$Var\_name)}; due to the
potential runtime and memory commitment, we strongly recommend that
users exercise caution if attempting this.

\hypertarget{the-icon-class-and-network-s3-generic}{%
\subsection{The ICON class and network S3
generic}\label{the-icon-class-and-network-s3-generic}}

Looking at the structure of the complex networks with
\texttt{utils::str} shows that \texttt{ICON} complex networks all have
two classes: \texttt{ICON} and \texttt{data.frame}. The latter provides
a suitable container for edge list objects with potential edge
attributes in rectangular format. The former, an S3 class, benefits
users in two ways. First, it provides certain guarantees about object
format, i.e., an unmodified complex network object acquired via
\texttt{ICON} will have the \texttt{ICON} S3 class and is guaranteed to
be a data frame containing an edge list in which each row represents a
single edge, the first two columns specify nodes that define the
corresponding edge, and additional columns define edge attributes. This
standard format guarantee allows users, among other things, to generate
code for one \texttt{ICON} dataset with assurances that it will function
effectively for other \texttt{ICON} datasets. Second, the S3 class
allows users to take advantage of relevant S3 generics. \texttt{ICON}
specifically defines the \texttt{as.network.ICON} method for
compatibility with the \texttt{as.network} generic in the
\texttt{network} R package. This allows users to easily analyze and
visualize \texttt{ICON} datasets with the reliable \texttt{network} and
\texttt{ggnetwork} R packages, respectively. Section
\ref{networkusecase} will provide sample code that highlights the
benefit of this feature.

\hypertarget{use-cases}{%
\section{Use cases}\label{use-cases}}

Before starting with the use cases, the following code chunk will load
the appropriate libraries and download the sample dataset.

\begin{Shaded}
\begin{Highlighting}[]
\CommentTok{# load necessary libraries}
\KeywordTok{library}\NormalTok{(}\StringTok{"ICON"}\NormalTok{)}
\KeywordTok{library}\NormalTok{(}\StringTok{"network"}\NormalTok{)}
\KeywordTok{library}\NormalTok{(}\StringTok{"ggnetwork"}\NormalTok{)}
\KeywordTok{library}\NormalTok{(}\StringTok{"ggplot2"}\NormalTok{)}
\KeywordTok{library}\NormalTok{(}\StringTok{"igraph"}\NormalTok{)}

\CommentTok{# for reproducibility}
\KeywordTok{set.seed}\NormalTok{(}\DecValTok{42}\NormalTok{)}

\CommentTok{# download sample dataset}
\KeywordTok{get_data}\NormalTok{(}\StringTok{"seed_disperse_beehler"}\NormalTok{)}
\end{Highlighting}
\end{Shaded}

\begin{verbatim}
## DATASET(S) seed_disperse_beehler LOADED
\end{verbatim}

A quick exploration of \texttt{seed\_disperse\_beehler} will grant a
deeper understanding of the use cases. Primarily, we would like to
explore the third column - named \texttt{Frequency}. Due to the heavy
skew, we will use two consecutive logarithmic transformations to more
easily see the effects of coloring edges by the \texttt{Frequency} edge
attribute. The following code chunk produces histograms of
\texttt{seed\_disperse\_beehler\$Frequency} before and after this
transformation for comparison.

\begin{Shaded}
\begin{Highlighting}[]
\CommentTok{# plot a histogram w/o transformation (skewed, tough to see differences)}
\KeywordTok{ggplot}\NormalTok{(seed_disperse_beehler, }\KeywordTok{aes}\NormalTok{(}\DataTypeTok{x =}\NormalTok{ Frequency)) }\OperatorTok{+}
\StringTok{  }\KeywordTok{geom_histogram}\NormalTok{(}\DataTypeTok{bins =} \DecValTok{10}\NormalTok{, }\DataTypeTok{fill =} \StringTok{"white"}\NormalTok{, }\DataTypeTok{color =} \StringTok{"black"}\NormalTok{) }\OperatorTok{+}
\StringTok{  }\KeywordTok{theme_bw}\NormalTok{()}

\CommentTok{# plot a histogram w/ transformation (more spread out, differences easily seen)}
\KeywordTok{ggplot}\NormalTok{(seed_disperse_beehler, }\KeywordTok{aes}\NormalTok{(}\DataTypeTok{x =} \KeywordTok{log}\NormalTok{(}\KeywordTok{log}\NormalTok{(Frequency)))) }\OperatorTok{+}
\StringTok{  }\KeywordTok{geom_histogram}\NormalTok{(}\DataTypeTok{bins =} \DecValTok{10}\NormalTok{, }\DataTypeTok{fill =} \StringTok{"white"}\NormalTok{, }\DataTypeTok{color =} \StringTok{"black"}\NormalTok{) }\OperatorTok{+}
\StringTok{  }\KeywordTok{theme_bw}\NormalTok{()}
\end{Highlighting}
\end{Shaded}

\includegraphics{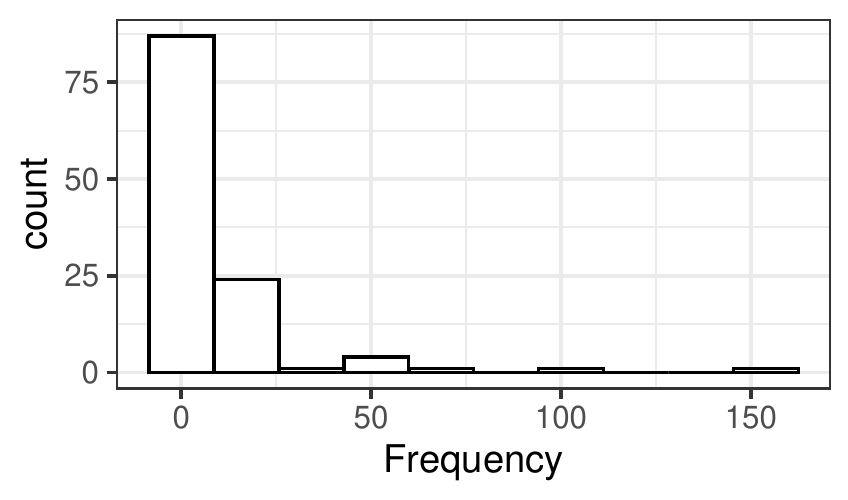}
\includegraphics{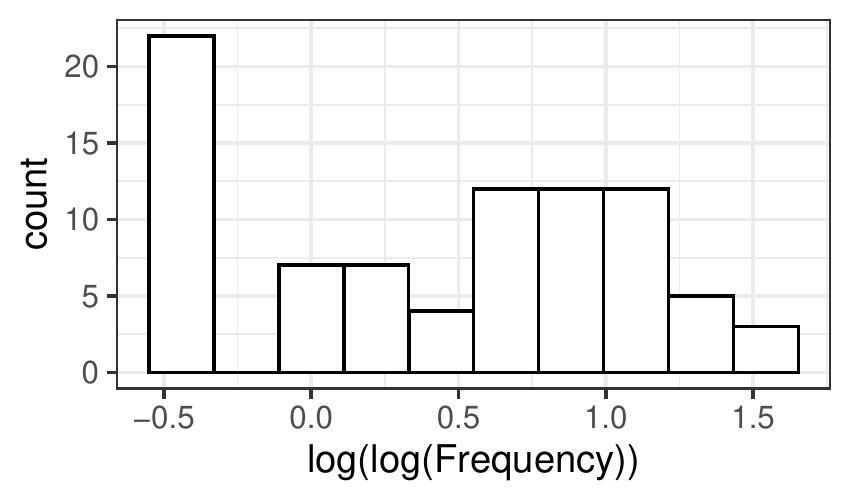}

\hypertarget{networkusecase}{%
\subsection{\texorpdfstring{With the \texttt{network} R
package}{With the network R package}}\label{networkusecase}}

Using the \texttt{seed\_disperse\_beehler} sample dataset, we first
coerce it to have class \texttt{network} with the
\texttt{as.network.ICON} method under the \texttt{as.network} S3
generic. This allows us to take advantage of the large set of tools
already built in the \href{http://www.statnet.org}{Statnet suite} of R
packages. Although we first use \texttt{ggnetwork} to rapidly visualize
the nodes and edges, we also show how to visualize edge attributes
toward the end of the code chunk.

\begin{Shaded}
\begin{Highlighting}[]
\CommentTok{# setup using as.network generic}
\NormalTok{coerced <-}\StringTok{ }\KeywordTok{as.network}\NormalTok{(seed_disperse_beehler)}

\CommentTok{# plot with ggnetwork}
\KeywordTok{ggplot}\NormalTok{(coerced, }\KeywordTok{aes}\NormalTok{(}\DataTypeTok{x =}\NormalTok{ x, }\DataTypeTok{y =}\NormalTok{ y, }\DataTypeTok{xend =}\NormalTok{ xend, }\DataTypeTok{yend =}\NormalTok{ yend)) }\OperatorTok{+}
\StringTok{  }\KeywordTok{geom_edges}\NormalTok{(}\DataTypeTok{alpha =} \FloatTok{0.25}\NormalTok{) }\OperatorTok{+}
\StringTok{  }\KeywordTok{geom_nodes}\NormalTok{() }\OperatorTok{+}
\StringTok{  }\KeywordTok{theme_blank}\NormalTok{()}
\end{Highlighting}
\end{Shaded}

\includegraphics{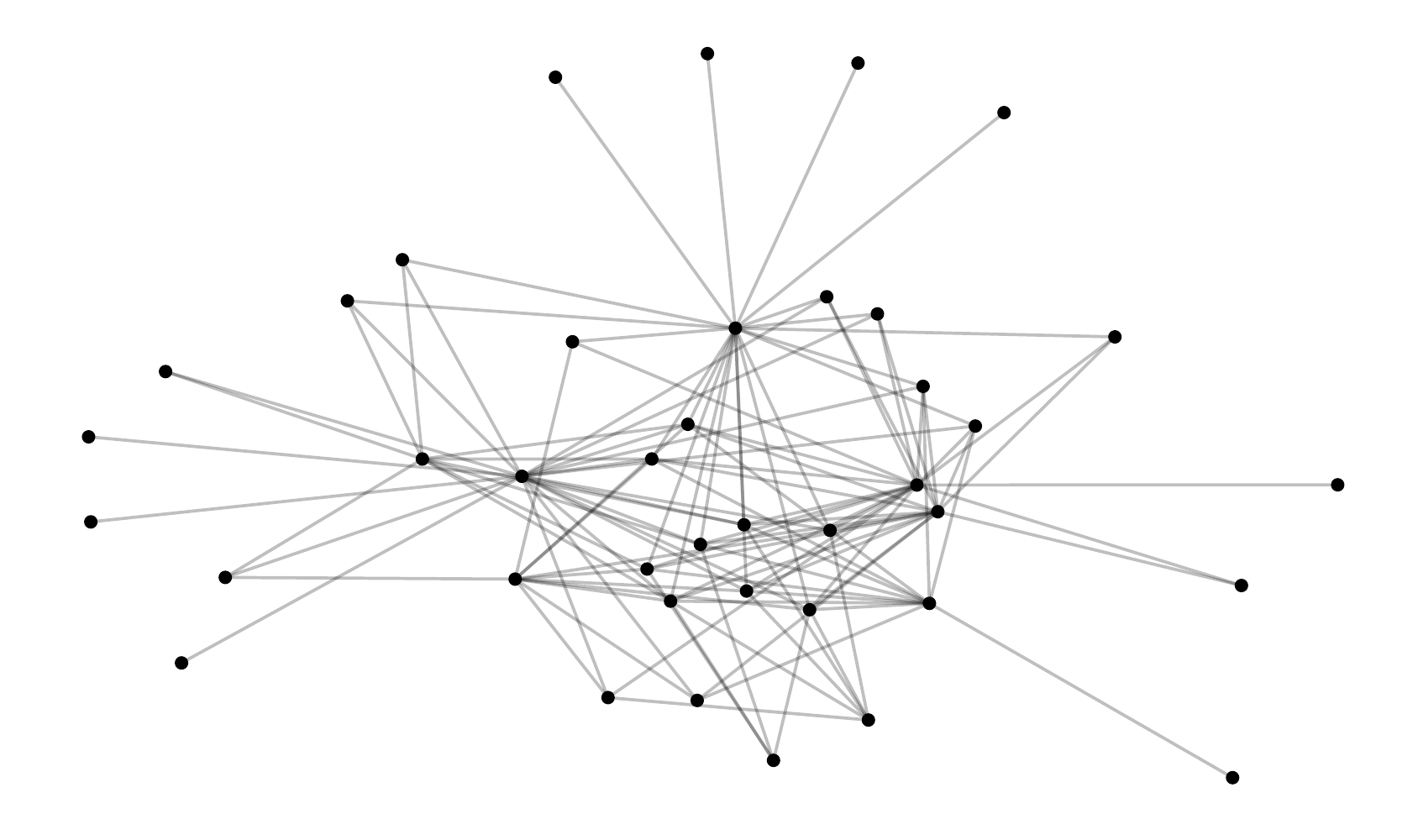}

\begin{Shaded}
\begin{Highlighting}[]
\CommentTok{# are there any edge attributes in `seed_disperse_beehler`?}
\CommentTok{#   YES, we have the "Frequency" edge attribute (see third column name)}
\KeywordTok{str}\NormalTok{(seed_disperse_beehler)}
\end{Highlighting}
\end{Shaded}

\begin{verbatim}
## Classes 'ICON' and 'data.frame': 119 obs. of  3 variables:
##  $ Bird     : chr  "1" "1" "1" "1" ...
##  $ Plant    : chr  "V4" "V6" "V8" "V11" ...
##  $ Frequency: num  1 1 1 25 19 1 1 8 1 1 ...
\end{verbatim}

\begin{Shaded}
\begin{Highlighting}[]
\CommentTok{# is this edge attribute also present in the coerced network?}
\CommentTok{#   YES, let's plot it in the next network visualization (see end of output)}
\KeywordTok{print}\NormalTok{(coerced)}
\end{Highlighting}
\end{Shaded}

\begin{verbatim}
##  Network attributes:
##   vertices = 40 
##   directed = FALSE 
##   hyper = FALSE 
##   loops = FALSE 
##   multiple = FALSE 
##   bipartite = FALSE 
##   total edges= 119 
##     missing edges= 0 
##     non-missing edges= 119 
## 
##  Vertex attribute names: 
##     vertex.names 
## 
##  Edge attribute names: 
##     Frequency
\end{verbatim}

\begin{Shaded}
\begin{Highlighting}[]
\CommentTok{# plot with log(log(Frequency)) as an edge attribute (edge color)}
\KeywordTok{ggplot}\NormalTok{(coerced, }\KeywordTok{aes}\NormalTok{(}\DataTypeTok{x =}\NormalTok{ x, }\DataTypeTok{y =}\NormalTok{ y, }\DataTypeTok{xend =}\NormalTok{ xend, }\DataTypeTok{yend =}\NormalTok{ yend)) }\OperatorTok{+}
\StringTok{  }\KeywordTok{geom_edges}\NormalTok{(}\KeywordTok{aes}\NormalTok{(}\DataTypeTok{color =} \KeywordTok{log}\NormalTok{(}\KeywordTok{log}\NormalTok{(Frequency)))) }\OperatorTok{+}
\StringTok{  }\KeywordTok{geom_nodes}\NormalTok{() }\OperatorTok{+}
\StringTok{  }\KeywordTok{theme_blank}\NormalTok{()}
\end{Highlighting}
\end{Shaded}

\includegraphics{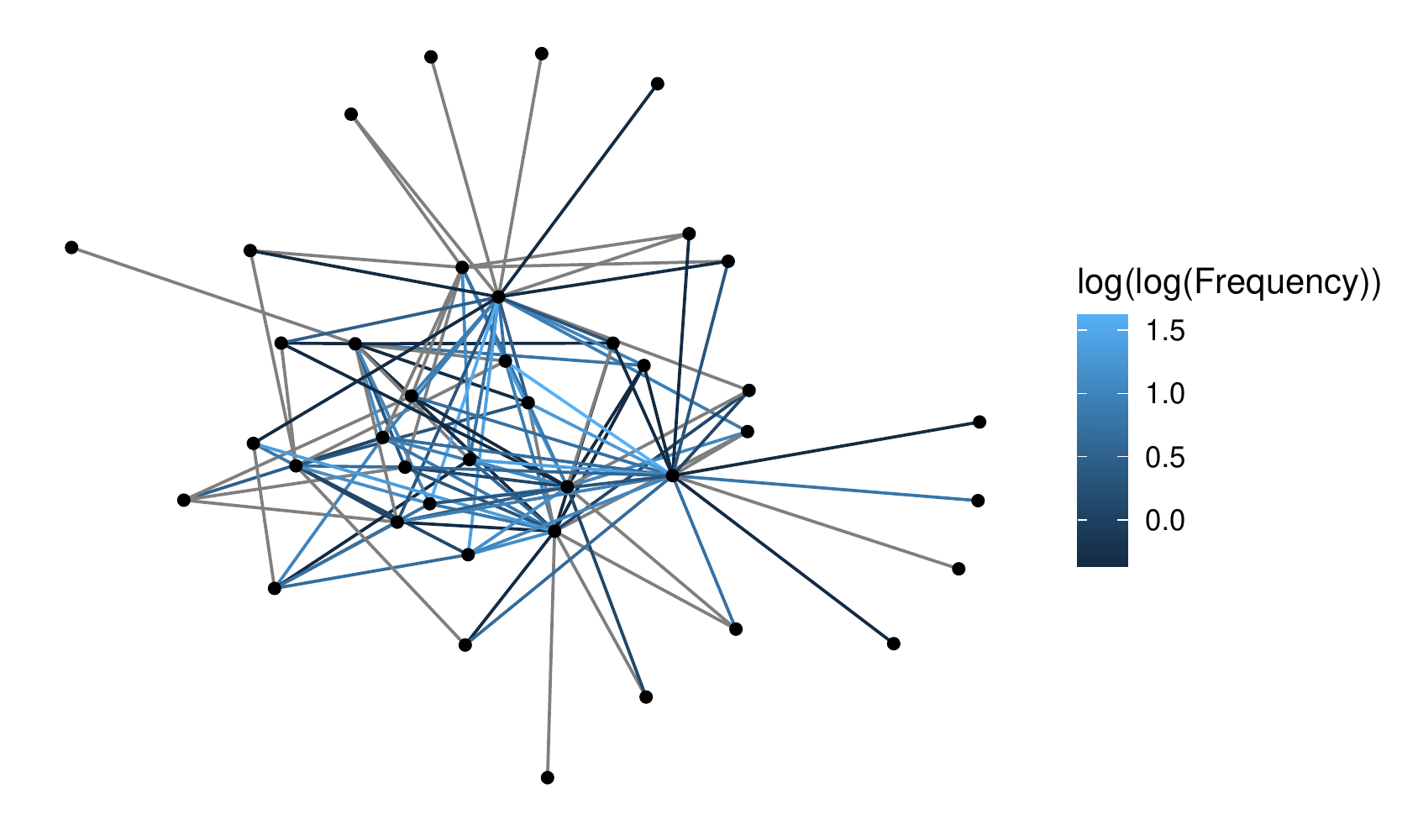}

Of course, even with the edge attribute, we are only scraping the
surface of the visualization capability provided. More details can be
found in the documentation of the appropriate packages.

\hypertarget{with-the-igraph-r-package}{%
\subsection{\texorpdfstring{With the \texttt{igraph} R
package}{With the igraph R package}}\label{with-the-igraph-r-package}}

The \texttt{igraph} package provides a rich set of network analysis and
visualization tools. As a consequence, the installed package is large.
Making \texttt{ICON} strongly dependent on \texttt{igraph} by building
an \texttt{as.igraph.ICON} method for the \texttt{as.igraph} S3 generic
would require all \texttt{ICON} users to also install \texttt{igraph}.
To avoid this, \texttt{ICON} does not strongly depend on
\texttt{igraph}, however, the following code chunk and the
\href{https://github.com/rrrlw/ICON/blob/master/README.md}{GitHub
README} demonstrate how to use \texttt{ICON} in conjunction with
\texttt{igraph} to emphasize that, despite the size limitation that
excluded it from being an \texttt{ICON} dependency, it is a powerful
tool. Instead of coercing \texttt{seed\_disperse\_beehler} into an
object with class \texttt{igraph}, we transfer the information in
\texttt{seed\_disperse\_beehler} to a new \texttt{igraph} object with
\texttt{igraph::graph\_from\_edgelist} and then visualize the network
using the \texttt{igraph::plot.igraph} method for the
\texttt{base::plot} S3 generic.

\begin{Shaded}
\begin{Highlighting}[]
\CommentTok{# plot network using igraph}
\NormalTok{my_graph <-}\StringTok{ }\KeywordTok{graph_from_edgelist}\NormalTok{(}\KeywordTok{as.matrix}\NormalTok{(seed_disperse_beehler[, }\DecValTok{1}\OperatorTok{:}\DecValTok{2}\NormalTok{]),}
                                \DataTypeTok{directed =} \OtherTok{FALSE}\NormalTok{)}
\KeywordTok{plot}\NormalTok{(my_graph,}
     \DataTypeTok{vertex.label =} \OtherTok{NA}\NormalTok{, }\DataTypeTok{vertex.size =} \DecValTok{5}\NormalTok{)}
\end{Highlighting}
\end{Shaded}

\includegraphics{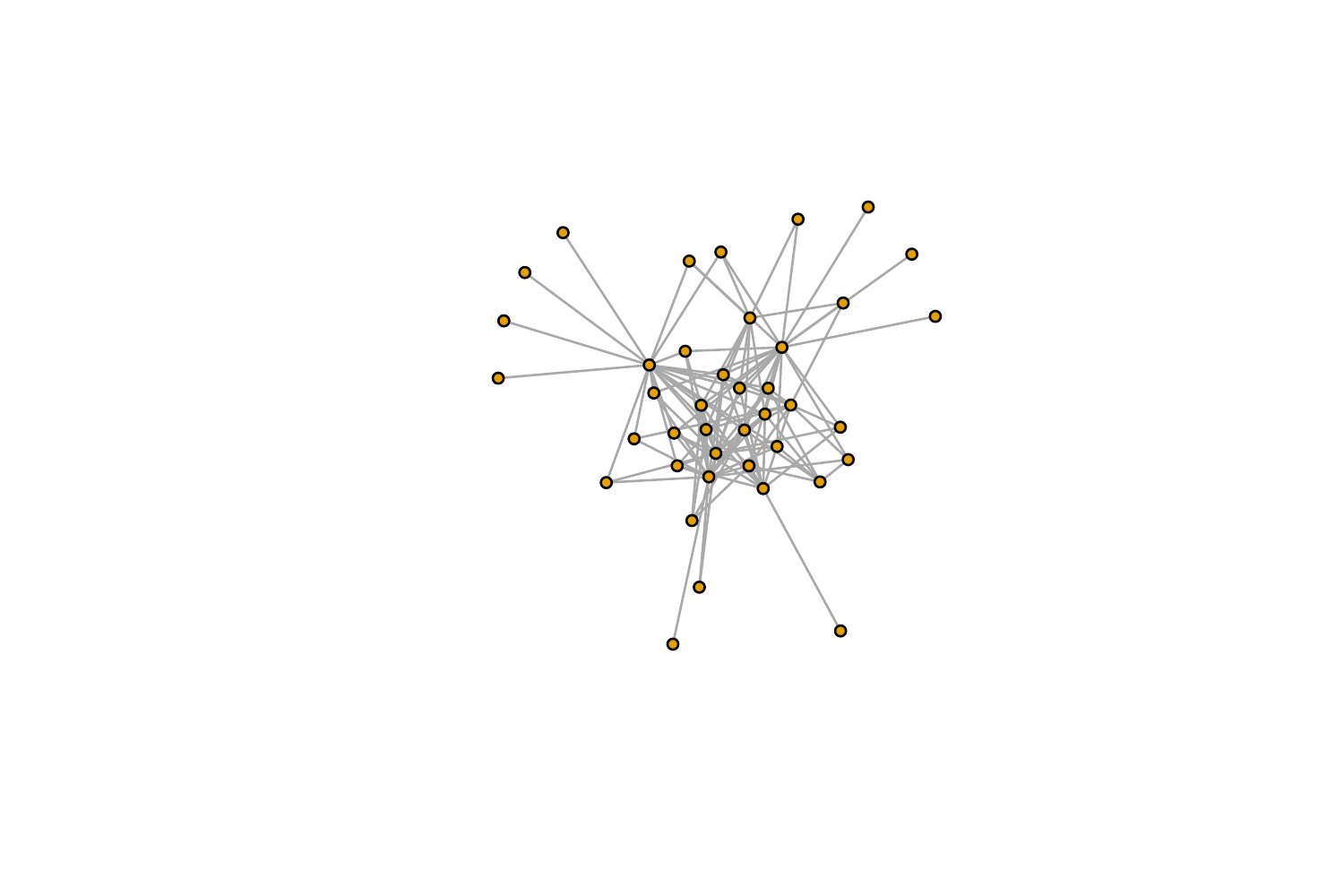}

As was the case with the previous use case, we have only scratched the
surface of visualization possibilities. More details can be found in the
\texttt{igraph} package documentation.

\hypertarget{discussionconclusions}{%
\section{Discussion/Conclusions}\label{discussionconclusions}}

We have introduced the \texttt{ICON} R package, explained its potential use as a network corpus, and demonstrated its compatibility with existing complex network software. 
With time, we hope that \texttt{ICON}'s corpus will grow and encourage users to contribute complex network datasets by following steps in the package's \href{https://github.com/rrrlw/ICON/blob/master/CONTRIBUTING.md}{contributing
guidelines} and adhering to the \href{https://github.com/rrrlw/ICON/blob/master/CONTRIBUTING.md}{code of conduct},\textsuperscript{15} both of which can be found on the package's GitHub repository. More details about the \texttt{ICON} R package can be found at the package website (\url{https://rrrlw.github.io/ICON/}), GitHub repository (\url{https://github.com/rrrlw/ICON}), and CRAN page (\url{https://CRAN.R-project.org/package=ICON}).

\hypertarget{acknowledgments}{%
\section{Acknowledgments}\label{acknowledgments}}

The authors thank Aaron Clauset, PhD and the members of his research group at the University of Colorado Boulder for advice and for their tireless efforts in creating the Colorado Index of Complex Networks.

\hypertarget{references}{%
\section*{References}\label{references}}
\addcontentsline{toc}{section}{References}

\hypertarget{refs}{}
\leavevmode\hypertarget{ref-corpus1}{}%
1. Ghasemian A, Hosseinmardi H, Clauset A. Evaluating overfit and
underfit in models of network community structure. IEEE Transactions on
Knowledge and Data Engineering 2020;32:1722-1735.

\leavevmode\hypertarget{ref-corpus2}{}%
2. Broido AD, Clauset A. Scale-free networks are rare. Nature
Communications 2019;10:1017.

\leavevmode\hypertarget{ref-icon}{}%
3. Clauset A, Tucker E, Sainz M. The Colorado Index of Complex
Networks., 2016: Available at https://icon.colorado.edu.

\leavevmode\hypertarget{ref-rlang}{}%
4. R Core Team. R: A Language and Environment for Statistical Computing.
Vienna, Austria: R Foundation for Statistical Computing, 2020: Available
at https://www.R-project.org.

\leavevmode\hypertarget{ref-network-pkg}{}%
5. Butts C. network: A package for managing relational data in R.
Journal of Statistical Software 2008;24:1-36.

\leavevmode\hypertarget{ref-covr-pkg}{}%
6. Hester J. covr: Test Coverage for Packages., 2020: Available at
https://CRAN.R-project.org/package=covr.

\leavevmode\hypertarget{ref-ggnetwork-pkg}{}%
7. Briatte F. ggnetwork: Geometries to Plot Networks with 'ggplot2'.,
2020: Available at https://CRAN.R-project.org/package=ggnetwork.

\leavevmode\hypertarget{ref-ggplot2-pkg}{}%
8. Wickham H. ggplot2: Elegant Graphics for Data Analysis.
Springer-Verlag: New York, NY, 2016.

\leavevmode\hypertarget{ref-knitr-pkg}{}%
9. Xie Y. Dynamic Documents with R and Knitr. 2nd ed. Boca Raton,
Florida: Chapman; Hall/CRC, 2015.

\leavevmode\hypertarget{ref-testthat-pkg}{}%
10. Wickham H. testthat: get started with testing. The R Journal
2011;3:5-10.

\leavevmode\hypertarget{ref-rmarkdown-pkg}{}%
11. Xie Y, Allaire JJ, Grolemund G. R Markdown: The Definitive Guide.
Boca Raton, Florida: Chapman; Hall/CRC, 2018.
\url{https://bookdown.org/yihui/rmarkdown}.

\leavevmode\hypertarget{ref-devtools-pkg}{}%
12. Wickham H, Hester J, Chang W. devtools: Tools to Make Developing R
Packages Easier., 2020: Available at
https://CRAN.R-project.org/package=devtools.

\leavevmode\hypertarget{ref-remotes-pkg}{}%
13. Hester J, Csárdi G, Wickham H, et al. remotes: R Package
Installation from Remote Repositories, Including 'Github'., 2020:
Available at https://CRAN.R-project.org/package=remotes.

\leavevmode\hypertarget{ref-roxygen2-pkg}{}%
14. Wickham H, Danenberg P, Csárdi G, et al. roxygen2: in-Line
Documentation for R., 2020: Available at
https://CRAN.R-project.org/package=roxygen2.

\leavevmode\hypertarget{ref-covenant}{}%
15. Contributor Covenant: A Code of Conduct for Open Source Projects,
2014: Available at https://www.contributor-covenant.org.

\leavevmode\hypertarget{ref-rpkg-book}{}%
16. Wickham H. R Packages. O'Reilly Media: Sebastopol, CA, 2015.


\end{document}